\documentclass[12pt]{iopart}
\usepackage[dvips]{graphicx}
\newcommand{\be}{\begin{equation}}
\newcommand{\ee}{\end{equation}}
\begin{document}
\title{Holography and holographic dark energy model}
\author{Yungui Gong}
\address{College of Electronic
Engineering, Chongqing University of Posts and Telecommunications,
Chongqing 400065, China} \ead{gongyg@cqupt.edu.cn}
\author{Yuan-Zhong Zhang}
\address{Institute of Theoretical Physics, Chinese Academy of Sciences,
P.O. Box 2735, Beijing 100080, China}
\begin{abstract}
The holographic principle is used to discuss the holographic dark energy model.
We find that the Bekenstein-Hawking entropy bound is far from saturation under certain conditions.
A more general constraint on the parameter of the holographic dark energy model
is also derived.
\end{abstract}
 \maketitle
\parindent=4ex

The studies of black hole thermodynamics tell us that
the total entropy of matter inside a black hole cannot be greater
than the Bekenstein-Hawking entropy, which is one quarter of the
area of the event horizon of the black hole measured in Planck
unit. Bekenstein proposed a universal entropy bound $S\le
2\pi ER$ for a weakly self-gravitating physical system
with total energy $E$ and size $R$ in an asymptotically flat
four dimensional spacetime \cite{bekenstein} in 1981. 't Hooft and
Susskind later introduced the Bekenstein-Hawking entropy as the
holographic entropy bound \cite{holography}. Based on the earlier
works, Bousso proposed the covariant entropy bound in
\cite{bousso}. The conjectured Anti-de-Sitter/Conformal
Field Theory duality provided the evidence of the existence
of a holographic principle in quantum gravity although we are far
from understanding quantum gravity \cite{adscft}. According to the
holographic principle, under certain conditions all the
information about a physical system inside a spatial region is encoded
in its boundary instead of its volume. Fischler and Susskind (FS)
discussed the application of the holographic principle to
cosmology by considering our universe inside a particle horizon
\cite{fischler}. They found that the holographic bound was
violated for a closed universe. To solve this problem, Bak and Rey
then replaced the particle horizon by the apparent horizon
\cite{bak}, and Rama found that the holographic bound
with the particle horizon could be satisfied in a closed universe
by adding negative pressure matter \cite{rama}.
There are other modifications to the original FS version of holographic principle \cite{mod}. The
holographic principle was also used to discuss the number of
e-foldings of inflation \cite{bound} and constrain dark energy models
\cite{wang}. The discussion of
holographic principle in the context of Brans-Dicke theory was
first considered in \cite{gong00}.

The type Ia supernova observations suggest that the
Universe is dominated by dark energy with negative pressure which
provides the dynamical mechanism of the accelerating expansion of
the Universe \cite{sp99,agr98}. The simplest candidate of dark
energy is the cosmological constant. However, the unusual small
value of the cosmological constant is a big challenge to
theoretical physicists. The idea of holography may be used to
solve the cosmological constant problem \cite{hldark1,lvdark}.
Cohen, Kaplan and Nelson proposed that for any state with energy $E$ in the
Hilbert space, the corresponding Schwarzschild
radius $R_s\sim E$ is less than the infrared (IR) cutoff $L$
\cite{hldark1}. Under this assumption, a relationship between the
ultraviolet (UV) cutoff $\rho_D^{1/4}$ and the IR cutoff is derived, i.e.,
$8\pi GL^3\rho_D/3\sim L$ \cite{hldark1}. So the holographic
dark energy density is
\begin{equation}
\label{hldark} \rho_D= \frac{3c^2\,d^2}{8\pi G L^2},
\end{equation}
where $c$ is the speed of light and $d$ is a constant of the order
of unity. Hsu found that the holographic dark energy
model based on the Hubble scale as the IR cutoff won't give an
accelerating universe \cite{hldark2}. This does not mean that the form
$\rho_D\sim H^2$ won't never work for dark energy model building.
The model $\rho_D=\Lambda+3c^2d^2H^2/(8\pi G)$ derived from the
re-normalization group models of the cosmological constant can explain the
accelerating expansion of the Universe \cite{hubble}. Ito discovered a viable
holographic dark energy model by using the Hubble scale as the IR
cutoff with the use of non-minimal coupling to scalar field
\cite{ito}. More recently, a dark energy model $\rho_D\sim H^2$ with the interaction
between dark energy and dark matter was proposed \cite{pavon}.
In \cite{hldark3}, Li showed
that the holographic dark energy model based on the
particle horizon as the IR cutoff won't give an accelerating
universe either. However, Li found that the holographic dark
energy model based on the event horizon as the IR cutoff works
\cite{hldark3,huang2}. The model was also found to be consistent
with current observations \cite{huang1}. The holographic dark energy model in the framework of
Brans-Dicke theory was discussed in \cite{gong04}. Some
speculations about the deep reasons of the holographic dark energy
were considered by several authors \cite{origin}. The holographic dark energy model was also
discussed in \cite{gong05,hldark4}.

For a fluid with constant equation of state $p=w\rho$, the second law
of thermodynamics tells us that a relationship $\rho\sim\sigma^{1+w}$ between
the entropy density $\sigma$ and the energy density $\rho$. If we apply the
Bekenstein-Hawking entropy bound, then we will get an upper bound on $\rho$. If we
apply the energy bound $R_s\le L$, then we will get an upper bound on entropy.
The consistency between the entropy bound derived from the energy bound and the holographic
bound needs to be checked. Furthermore, if we use the Bekenstein entropy bound, a lower
bound on $\rho$ is obtained. Therefore, a detailed discussion about the effects of those arguments
on the holographic dark energy model is needed.
In this paper, we first review the holographic dark energy model \cite{hldark3,huang2,gong05},
then we apply the second law of thermodynamics and the holographic principle to discuss the
consistency of those bounds. For a general perfect fluid, there is no relationship between $\sigma$ and
$\rho$, but the comoving entropy density is a constant. This fact is used to prove that the Bekenstein-Hawking
entropy bound is far from saturation for the holographic dark energy model.

We use the homogeneous and isotropic Friedmann-Robertson-Walker (FRW) metric
\begin{equation}
\label{rwcosm} ds^2=-c^2dt^2+a^2(t)\left[{dr^2\over
1-k\,r^2}+r^2\,d\Omega\right].
\end{equation}
For a null geodesic, we have \be \label{line}
\int_{t_1}^{t_0}{c\,dt\over a(t)}=\int_0^{r_1}{dr\over
\sqrt{1-kr^2}}\equiv f(r_1), \ee where
\begin{equation*}
f(r_1)=\left\{\begin{array}{ll}
\sin^{-1}(\sqrt{|k|}\,r_1)/\sqrt{|k|},\ \ \ \ \ \ &k=1,\\
r_1,&k=0,\\
\sinh^{-1}(\sqrt{|k|}\,r_1)/\sqrt{|k|},&k=-1.
\end{array}\right.
\end{equation*}
With both an ordinary pressureless dust matter and the holographic
dark energy as sources, the Friedmann equations are
\be
\label{cos1} H^2+{kc^2\over a^2}={8\pi
G\over 3}(\rho_m+\rho_r+\rho_D),
\ee
\be
\label{cos2} \dot{\rho_D}+3H(\rho_D+p_D)=0,
\ee
where the Hubble parameter $H=\dot{a}/a$, dot means
derivative with respect to time, the matter density
$\rho_m=\rho_{m0}(1/a)^3$, the radiation density
$\rho_r=\rho_{r0}(1/a)^4$, a subscript 0 means the value of the
variable at present time and we set $a_0=1$. Define the proper
event horizon as
\be\label{reh} R_{eh}(t)=a(t)\int^\infty_t
\frac{cdt}{a(t)}=a(t)\int^\infty_{a(t)}
\frac{cd\tilde{a}}{\tilde{a}^2H}.\ee
If we choose the event horizon as the IR cutoff, then with the help of Eq. (\ref{line}),
we have
\be\label{leh} L=a(t)r_1=\frac{a(t) {\rm
sinn}[\sqrt{|k|}\,R_{eh}(t)/a(t)]}{\sqrt{|k|}}, \ee
where ${\rm sinn}(x)=\sin(x)$\{$x$, $\sinh(x)$\} if $k=1$\{$0$, $-1$\}
respectively. For the flat case $k=0$, we recover $L=R_{eh}$.

Let us rewrite Eq. (\ref{cos1}) as \be \label{frweq}
\Omega_m+\Omega_r+\Omega_D=1+\Omega_k, \ee where
$\Omega_m=\rho_m/\rho_{cr}=\Omega_{m0}H^2_0/(H^2a^3)$,
$\Omega_r=\rho_r/\rho_{cr}=\Omega_{r0}H^2_0/(H^2a^4)$,
$\Omega_D=\rho_D/\rho_{cr}=d^2c^2/(L^2H^2)$,
$\Omega_k=kc^2/(a^2H^2)=\Omega_{k0}H^2_0/(a^2H^2)$ and $\rho_{cr}=3H^2/(8\pi G)$. Since
$$\frac{\Omega_k}{\Omega_m}=a\frac{\Omega_{k0}}{\Omega_{m0}}=a\gamma,$$
where $\gamma=\Omega_{k0}/\Omega_{m0}$, and
$$\frac{\Omega_r}{\Omega_m}=\frac{\Omega_{r0}}{a\Omega_{m0}}=\frac{\beta}{a},$$
where $\beta=\Omega_{r0}/\Omega_{m0}=1/(1+z_{eq})$ and the matter
radiation equality redshift $z_{eq}=3233$ \cite{wmap}, so
\be
\label{wmeq}
\Omega_m=\frac{\Omega_{m0}H^2_0}{H^2a^3}=\frac{a(1-\Omega_D)}{\beta+a-a^2\gamma}.
\ee
From the above equation, we get
\be \label{wmeq1}
\frac{1}{aH}=\frac{a}{H_0}\sqrt{\frac{1-\Omega_D}{\Omega_{m0}(\beta+a-a^2\gamma)}}.
\ee
Combining Eqs. (\ref{leh}) and (\ref{wmeq1}), we get
\begin{eqnarray} \label{wleq}
\sqrt{|k|}\frac{R_{eh}}{a}&=&{\rm sinn}^{-1}
\left[d\sqrt{|\gamma|}\sqrt{\frac{a^2(1-\Omega_D)}{\Omega_D(\beta+a-a^2\gamma)}}\,\right]\nonumber\\
&=&{\rm sinn}^{-1}(d\,\sqrt{|\Omega_k|/\Omega_D}).
\end{eqnarray}
If $\Omega_k>0$, then we require $d\le
\sqrt{\Omega_D/\Omega_k}$. Take derivative with respect to
$a$ on both sides of the above Eq. (\ref{wleq}) and use the
redshift $z=1/a-1$ as the variable, we finally get the following
differential equation by using Eqs. (\ref{reh}) and (\ref{wmeq1})
\begin{eqnarray} \label{wldeq}
\frac{d\Omega_D}{dz}&=&-\frac{2\Omega_D^{3/2}(1-\Omega_D)}{d(1+z)}
\sqrt{1-\frac{d^2\gamma(1-\Omega_D)}{\Omega_D[\beta(1+z)^2+1+z-\gamma]}}\nonumber\\
&&-\frac{\Omega_D(1-\Omega_D)[1+2\beta(1+z)]}{\beta(1+z)^2+1+z-\gamma}.
\end{eqnarray}

Combining Eqs. (\ref{hldark}),
(\ref{cos2})-(\ref{leh}) and (\ref{wleq}), we get the
dark energy equation of state
\begin{eqnarray} \label{ehol}
w_D&=&-\frac{1}{3}\left[1+\frac{2}{d}\sqrt{\Omega_D}{\rm
cosn}(\sqrt{|k|}\,R_{eh}/a)\right]\nonumber\\
&=&-\frac{1}{3}\left[1+\frac{2}{d}\sqrt{\Omega_D-d^2\Omega_k}\right],
\end{eqnarray}
where ${\rm cosn}(x)=\cos(x)$\{$1$, $\cosh(x)$\} if $k=1$\{$0$, $-1$\}
respectively. It is obvious that $w_D\le -1/3$, so the the Universe is expanding with acceleration
when we choose the event horizon as the IR cutoff. The acceleration is also encoded
in the existence of the event horizon.

From the second law of thermodynamics, we have
\be
\label{2nd}
TdS(T,V)=Td[\sigma(T,V) V]=d(\rho(T)V)+p(T)dV,
\ee
where $\sigma(T,V)$ is the entropy density. The integrability condition
$\partial^2 S/\partial V\partial T=\partial^2 S/\partial T\partial V$ gives us that
\be
\label{ptrel}
\frac{dp}{dT}=\frac{\rho+p}{T}.
\ee
Substitute Eq. (\ref{ptrel}) into Eq. (\ref{2nd}), we get
\be
\label{ptrel1}
\sigma=\frac{\rho+p}{T}.
\ee
Combining Eqs. (\ref{2nd}), (\ref{ptrel}) and (\ref{ptrel1}), we get
\be
\label{ptrel2}
d\sigma=\frac{d\rho}{T}.
\ee
Combining Eqs. (\ref{ptrel1}), (\ref{ptrel2}) and the energy conservation equation (\ref{cos2}), we get
\be
\label{rel4}
\sigma a^3={\rm constant}.
\ee
Therefore, the second law of thermodynamics tells us that the comoving entropy density
$\sigma'=\sigma a^3$ is a constant. For a fluid with constant equation of state
$p=w\rho$, Eqs. (\ref{ptrel1}) and (\ref{ptrel2})
give us the familiar relation $\sigma\sim \rho^{1/(1+w)}$.

Now let us check if the holographic dark energy model satisfies the holographic principle.
The total area of the system is
\be
\label{area}
A=4\pi L^2=4\pi a^2 d^2\frac{|\Omega_k|}{|k|\Omega_D}.
\ee
So
\begin{eqnarray}
\label{areadot}
\dot{A}=8\pi L\dot{L}&=&8\pi L\{LH-c\,{\rm
cosn}[\sqrt{|k|}\,R_{eh}(t)/a(t)]\}\nonumber\\
&=&8\pi Lc\left(\frac{d}{\sqrt{\Omega_D}}-\sqrt{1-\frac{d^2\gamma(1-\Omega_D)}
{\Omega_D[\beta(1+z)^2+1+z-\gamma]}}\right) \nonumber\\&\ge&0,
\end{eqnarray}
if
\begin{equation}
 \label{dlbnd}
 d^2\ge \frac{\Omega_D[\beta(1+z)^2+1+z-\gamma]}
 {\beta(1+z)^2+1+z-\gamma\Omega_D}=\frac{\Omega_D}{1+\Omega_k}.
 \end{equation}
 For the spatially flat universe, we recover $d^2\ge
\Omega_D$ \cite{huang2}. The constraint (\ref{dlbnd}) was derived in
\cite{huang2,gong05}.
Substitute the above inequality (\ref{dlbnd})
into Eq. (\ref{ehol}), we find that
$w_D\ge -1$. Therefore there is no phantom behavior for
the holographic dark energy model.

The total comoving volume of the system is
\begin{eqnarray}
\label{vlm}
V&=&\frac{\pi}{k}\frac{2\sqrt{|k|}R_{eh}/a-{\rm sinn}(2\sqrt{|k|}R_{eh}/a)}{\sqrt{|k|}},\nonumber\\
&=&\frac{\pi}{k}\frac{2{\rm sinn}^{-1}(d\sqrt{|\Omega_k|/\Omega_D})-
2d\sqrt{|\Omega_k|/\Omega_D}\sqrt{1-d^2\Omega_k/\Omega_D}}{\sqrt{|k|}}.
\end{eqnarray}
So
\be
\label{vlmt}
\dot{V}=\frac{-4\pi c}{k a}\frac{d^2\Omega_k}{\Omega_D}=\frac{-4\pi c}{a^3}L^2\le 0.
\ee
Therefore, the comoving volume shrinks when time goes by. The total entropy of the holographic
dark energy decreases with time. However, for a spatially flat universe, the physical volume
keeps increasing when time goes by since
\be
\label{pvlmt}
\dot{V}_{phy}=\frac{d (a^3 V)}{dt}=3a^3 H V-4\pi c L^2=4\pi c L^2\left(\frac{d}{\sqrt{\Omega_D}}-1\right)\ge 0.
\ee

With the condition
(\ref{dlbnd}), the holographic principle can be easily satisfied in late
times if it is satisfied during the early times. In other words, the Bekenstein entropy
$S_{BH}=A/4G$ is far from saturation at late times. So any constraint from the holographic principle is
a weak constraint. Therefore, we consider the argument $R_s\le L$.
This is equivalent to say that the total energy in
a region of size $L$ should not exceed the
mass of a black hole with the same size. This energy argument was used to get the scaling
law for the upper bound of the entropy of a fluid with equation of state $p=w\rho$ \cite{sbnd},
$$S\sim L^{3-2/(1+w)}.$$
Since
\begin{equation}
\label{rs1}
R_s=\frac{2GM}{c^2}=2G\rho_D\left(\frac{V_{phy}}{c^2}\right)\le
L,
\end{equation}
so
\begin{equation}
\label{dubnd}
2{\rm sinn}^{-1}\left(d\sqrt{\frac{|\Omega_k|}{\Omega_D}}\right)
-2d\sqrt{\frac{|\Omega_k|}{\Omega_D}}\sqrt{1-d^2\frac{\Omega_k}{\Omega_D}}
\le \frac{4d}{3}\frac{k}{|k|}\left(\sqrt{\frac{|\Omega_k|}{\Omega_D}}\right)^3.
\end{equation}
For a spatially flat universe $k=0$, the above inequality gives $d^2\le 1$. Therefore, the
parameter $d$ must satisfy two constraint equations (\ref{dlbnd}) and (\ref{dubnd}). Because of the
constraint equation (\ref{dlbnd}), the holographic dark energy model has no phantom-like behavior and
the Bekenstein-Hawking entropy bound is far from saturation at late times.
For a spatially curved universe, the constraint
equation (\ref{dubnd}) by using the argument $R_s\le L$ is also derived. In conclusion, the energy
bound $R_s\le L$ gives the constraint Eq. (\ref{dubnd}) on the parameters in the holographic dark energy
model, the Bekenstein-Hawking entropy bound can be easily satisfied for the holographic dark energy model.

\ack
Y. Gong's work is supported by
NNSFC under grant No. 10447008, SRF for ROCS, State Education Ministry and
CQUPT under Grant No. A2004-05.
Y.Z. Zhang's work is in part supported by NNSFC
under Grant No. 90403032 and also by National Basic Research
Program of China under Grant No. 2003CB716300.

\end{document}